\documentclass{mem}
\usepackage{natbib}\usepackage{txfonts}\usepackage{balance}
\usepackage{graphicx}
\usepackage[a4paper]{hyperref}
\idline{75}{282}
\begin{document}
\def\teff{$T\rm_{eff }$}
\def\kms{$\mathrm {km s}^{-1}$}

\title{
Cosmological Models and the Brightness Profile of Distant Galaxies}

   \subtitle{}

\author{
I.\, Olivares-Salaverri\inst{1} 
\and Marcelo B.\, Ribeiro\inst{2}
          }

% \offprints{I.\ Olivares-Salaverri}

\institute{
Valongo Observatory,
Federal University of Rio de Janeiro-UFRJ,
Ladeira Pedro Ant\^{o}nio 43,
CEP: 20.080-090, Rio de Janeiro, RJ, Brazil;
\email{iker@astro.ufrj.br}
\and
Physics Institute, 
Federal University of Rio de Janeiro-UFRJ,
 CxP 68532, CEP 21945-970, Rio de Janeiro, RJ,
 Brazil; \email{mbr@if.ufrj.br}
}

\authorrunning{Olivares-Salaverri \& Ribeiro}

\titlerunning{Cosmology and Galactic Brightness Profile}

\abstract{
The aim of this project is to determine the consistency of an assumed
cosmological model by means of a detailed analysis of the brightness
profiles of distant galaxies. Starting from the theory developed by
Ellis and Perry (1979) connecting the angular diameter distance
obtained from a relativistic cosmological model and the detailed
photometry of galaxies, we assume the presently most accepted
cosmological model with non-zero cosmological constant and attempt to
predict the brightness profiles of galaxies of a given redshift. Then
this theoretical profile can be compared to observational data already
available for distant, that is, high redshift, galaxies. By comparing
these two curves we may reach conclusions about the observational
feasibility of the underlying cosmological model.
}
\maketitle{}

\section{Introduction}

The most basic goal of cosmology is to determine the spacetime geometry
and matter distribution of the Universe by means of astronomical
observations. Accomplishing this goal is not an easy or simple task, and
due to that, since the early days of modern cosmology several methods have
been advanced such that theory and observations are used to check one
another. Detailed analysis of the cosmic microwave background radiation,
galaxy number counts and supernova cosmology are just a few of the methods
employed nowadays in cosmology, deriving results that complement one another.
In this work we aim at discussing one of these methods, namely the
connection between galaxy brightness profiles and cosmological models.

Thirty years ago, Ellis and Perry (1979) advanced a very detailed
discussion where such a connection is explored. Their aim was to
determine the spacetime geometry of the universe by connecting the
angular diameter distance, also known as area distance, obtained from
a relativistic cosmological model, and the detailed photometry of
galaxies. They then discussed how the galaxy brightness profiles of
high redshift galaxies could be used to falsify cosmological models
as the angular diameter distance could be determined directly from
observations.

Nevertheless, to carry out this program to its full extent, one
would need detailed information about galaxy evolution.
Without a consistent theory on how galaxies evolve, it is presently
impossible to analyze cosmological observations without assuming a
cosmological model. In addition, brightness profiles are subject to
large observational errors, making it difficult to achieve
Ellis and Perry's aim of possibly using the angular diameter distance
determination to distinguish cosmological models.

This work is based on Ellis and Perry (1979) theory, although our
aim is more limited in the sense that we do not seek to determine
the underlying cosmological model by directly measuring the angular
diameter distance, but to assume the presently most
favored cosmology, deriving cosmological distances from it and seeking
to discuss the consistency between its predictions and detailed
observations of surface brightness of distant galaxies. Our goal is
to obtain a theoretical brightness profile by means of the assumed
cosmological model and compare it with its observational counterpart
at various redshift ranges, and for different galaxy morphologies.

The outline of the paper is as follow. In \S2 we introduce the
cosmological distances and their connections to astrophysical
observables. In \S3 we describe the parameters that determine the
surface brightness structure and in \S4 we discuss the criteria for
selecting galaxies, in view of the importance of evolutionary effects
in galactic surface brightness. 

\section{Cosmological Distances}

Let us consider that source and observer are at relative motion to
each other. From the point of view of the source, the light beams
that travel along future null geodesics define a solid angle
$d\Omega_{\scriptscriptstyle G}$ with the origin at the source and
have a transversal section area $d\sigma_{\scriptscriptstyle G}$ at
the observer. 

The flux  $F_{\scriptscriptstyle G}$ measured at the source considering
a 2-sphere $S$ lying in the locally Euclidean space-time centered
on the source is related to the source luminosity by, 
\begin{equation}
 L = \int_{S} F_{\scriptscriptstyle G} d\sigma_{\scriptscriptstyle G}
   = 4\pi F_{\scriptscriptstyle G}  
\end{equation}
assuming that it radiates with spherical symmetry and locally this
is a unit 2-sphere. If we consider now the flux $F_r$ radiated by
the source, but measured at the observer, the source
luminosity is 
\begin{equation}
 L = \int_{S} (1+z)^2 F_r d\sigma_{\scriptscriptstyle G}, 
\end{equation}
where the factor $(1+z)^2$ comes from \textit{area law} (Ellis 1971)
and $z$ is the redshift. This law establishes that the source luminosity
is independent from the observer. So these two equations are equal, and
we may write that,
\begin{equation}
L = \int_{S} F_{\scriptscriptstyle G} \; d\sigma_{\scriptscriptstyle G} =
\int_{S} (1+z)^2 F_r \; d\sigma_{\scriptscriptstyle G},
\end{equation}
\begin{equation}
(1+z)^2 F_r \; d\sigma_{\scriptscriptstyle G} = const =
F_{\scriptscriptstyle G} \; d\Omega_{\scriptscriptstyle G}
\label{e1}
\end{equation}
From the viewpoint of the source, we may now define the \textit{galaxy
area distance} $d_{\scriptscriptstyle G}$ as,
\begin{equation}
 d\sigma_{\scriptscriptstyle G} = {d_{\scriptscriptstyle G}}^2
 d\Omega_{\scriptscriptstyle G},
\end{equation}
which considering eq.\ (\ref{e1}), becomes,  
\begin{equation}
F_r = \frac{L}{4\pi}\frac{1}{(d_{\scriptscriptstyle G})^2(1+z)^2}.
\label{e2}
\end{equation}
The factor $(1+z)^2$ may be understood as arising from \textit{(i)}
the energy loss of each photon due to the redshift $z$, and \textit{(ii)}
the lower measured rate of arrival of photons due to the time dilation.
With eq.\ (\ref{e2}) it is not possible to make any physics since we cannot
measure the \textit{galaxy area distance} $d_{\scriptscriptstyle G}$.

Considering a bundle of null geodesics converging to the observer, that
is, light beams traveling from source to observer, they define a solid
angle $d\Omega_{\scriptscriptstyle A}$ with the origin at the observer and
have a transversal section area $d\sigma_{\scriptscriptstyle A}$ at the
source.  We may now define the \textit{angular diameter distance}
$d_{\scriptscriptstyle A}$ by
\begin{equation}
 d\sigma_{\scriptscriptstyle A} = {d_{\scriptscriptstyle A}}^2
 d\Omega_{\scriptscriptstyle A}.  
\end{equation}
The  \textit{reciprocity theorem}, due to Etherington (1933;
see also Ellis 1971, 2007) relates the $d_{\scriptscriptstyle G}$ and
$d_{\scriptscriptstyle A}$ by means of the following expression, 
\begin{equation}
{d_{\scriptscriptstyle G}}^2 = (1+z)^2 {d_{\scriptscriptstyle A}}^2.
\label{recip}
\end{equation}
This relation is purely geometric, valid for any cosmology and contains
information about spacetime curvature effects. Combining eqs.\ (\ref{e2})
and (\ref{recip}), it is possible to connect the flux received by the
observer and the \textit{angular diameter distance} by
\begin{equation}
 F_r = \frac{L}{4\pi {d_{\scriptscriptstyle G}}^2}\frac{1}{(1+z)^2} =
\frac{L}{4\pi {d_{\scriptscriptstyle A}}^2}\frac{1}{(1+z)^4}.
\end{equation}

\section{Connection with the surface photometry of cosmological sources}
Galaxies are objects that can be used to measure cosmological parameters
because they are located far enough in order to have significant spacetime
curvature effects. The flux emitted by these objects and received by the
observer depends on the surface brightness, which, by definition, is distant
independent, although it is redshift dependent (Ellis 2007). Based on the
reciprocity theorem, and bearing in mind that we actually observe in very
restricted wavelengths, it is possible to connect the emitted and received
\textit{specific surface brightness}, respectively denoted by $B_{e, \nu_e}$
and $B_{r, \nu_r}$, according to the following equation (Ellis and Perry 1979),
\begin{equation}
 B_{r,\nu_r}(\alpha ,z) = \frac{B_{e}(R,z)}{(1+z)^3} J[\nu_r (1+z),R, z].
\label{e4}
\end{equation}
Here $J$ is the \textit{spectral energy distribution} \textit{(SED)}, $R$ is
the \textit{intrinsic galactic radius}, $\nu_r$ and $\nu_e$ are respectively
the \textit{received} and \textit{emitted frequencies}, and $\alpha$ is defined
as the angle measured by the observer between the galactic center and its outer
luminous limit, as below (Ellis \& Perry 1979), 
\begin{equation}
 R = \alpha \; d_{\scriptstyle A} (z).
\end{equation}
Note that $d_{\scriptstyle A}$ is given by the assumed cosmological model. Our
aim is to compare the surface brightness observational data with its theoretically
derived results calculated by means of eq.\ (\ref{e4}) and reach conclusions about
the observational feasibility of assumed cosmological model.

To calculate the theoretical surface brightness, we have to assume some dependency
between the surface brightness and the intrinsic galactic radius.  Considering that
a fundamental assumption in observational cosmology is that homogeneous populations
of galaxies do exist, the structure and evolution of each member of such group of
galaxies will be essentially identical. This assumption implies that \textit{(i)}
the frequency dependence of the emitted galaxy radiation does not change across the
face of the galaxy, that is, it is $R$ independent, and \textit{(ii)} the radial
variation of the brightness is characterized by an amplitude $B_0$, which may
evolve with the redshift, i.e., $B_0(z)$, and a normalized radial functional form
does not evolve $f[R(z)/a(z)]$. So, the emitted surface brightness can be
characterized as (Ellis and Perry 1979),
\begin{equation}
 B_{e,\nu_e} (R,z) = B_{0}(z)J(\nu_e,z)f[R(z)/a(z)].
\label{e5}
\end{equation}
Now, let us define the parameter $\beta$ as being given by $\beta = R(z)/a(z)$,
where $a(z)$ is the scaling radius. The redshift dependence
in the parameters of the equation above is due to the galactic evolution. A
detailed study of the parameters of eq.\ (\ref{e5}) and their evolution is
fundamental to this work. Otherwise, we will not be able to infer if the
difference between the observational data and the modeled surface brightness
is due to the cosmological model or to a poor characterization of the brightness
structure and its evolution.

\subsection{Surface brightness profiles}
 
The function $f[R(z)/a(z)]$ characterize the shape of the surface brightness
distribution. There exist in the literature various different profiles. Some of
them are one parameter profiles, like \textit{Hubble} (1930), \textit{Hubble-Oemler}
and \textit{Abell-Mihalas} (1966), characterizing the galactic brightness
distribution quite well when the disk or bulge are dominant. They are given as,
\begin{equation}
 B_{\mathrm{H}, e, \nu_{e}} (R,z) = \frac{B_{0}(z)J(\nu_e,z)}{(1+\beta )^2};    
\end{equation}
\begin{equation}
 B_{\mathrm{HO}, e, \nu_{e}}(R,z) = \frac{B_{0}(z)J(\nu_e,z)e^{-R^2/R^2_t}}{(1+\beta )^2};
\end{equation}
\begin{eqnarray}
 B_{\mathrm{AM}, e, \nu_{e}}(R,z) & = & \frac{B_{0}(z)J(\nu_e,z)}{(1+\beta)^2}; \\
 & & (\beta \leq 21.4); \nonumber 
\end{eqnarray}
\begin{eqnarray}
 B_{\mathrm{AM}, e, \nu_{e}}(R,z) & = & \frac{22.4 B_{0}(z)J(\nu_e,z)}{(1+\beta)^2}; \\
 & & (\beta > 21.4). \nonumber
\end{eqnarray}
Other profiles like \textit{S\'{e}rsic} and \textit{core-S\'{e}rsic} use two or
more parameters, reproducing the galactic profile almost exactly (Trujillo et al.\
2004).
\begin{eqnarray}
 B_{\mathrm{S}, e, \nu_{e}}(R,z)= B_{eff}J(\nu_e,z) e^{ \left\{ - b_{n} \left[
 {\left( \frac{R}{R_{eff}} \right)}^{1/n} - 1 \right] \right\} }
\end{eqnarray}
\begin{eqnarray}
 B_{\mathrm{cS}, e, \nu_{e}}(R,z)&=&B_bJ(\nu_e,z) 2^{-\frac{\gamma}{\alpha}} {\left[1+{\left(\frac{R_b}{R}\right)}^{\alpha}
   \right]}^{\gamma/\alpha} \times \nonumber \\
   & \times & e^{\left[-b { \left( \frac{R^{\alpha}+R^{\alpha}_b}{R^{\alpha}_{eff}} \right) } ^{1/n\alpha}
  +b2^{1/\alpha n} {\left( \frac{R_b}{R_{eff}} \right)} ^{1/n}\right]},
\end{eqnarray}
where $B_{eff}$ is the surface brightness at the effective radius $R_{eff}$ that encloses
half of the total light, $B_b$ is the surface brightness at the core or break radius
$R_b$. $\gamma$ is the slope of the inner power-law region, $\alpha$ controls the sharpness
of the transition between the cusp and the outer S\'{e}rsic profile and $n$ is the shape
parameter of the outer S\'{e}rsic. The quantity $b$ is a function of the parameters
$\alpha$, $R_b/R_{eff}$, $\gamma$ and $n$. The parameter $b_n$ depends only on $n$. 

\section{Sample Selection Criteria}

To analyze only the effect of the cosmological model in the surface brightness and minimize
the effect of evolution, we assume that there exists a homogeneous class of objects whose
properties are similar in all redshifts, allowing us to carry out comparisons at different
values of z. Thus, galaxy sample selection follows this assumption. Choosing galaxies of
different morphologies, we must consider the following requirements: 
\begin{enumerate}
 \item[] {{\bf (1)} The existence of different morphological populations at different
 redshift values. Due to the Hubble sequence we know that not all type of galaxies exist
 in all epochs. Therefore, it seems reasonable to choose early-type galaxies because they
 exist at different redshift values and have a lower star formation rate which could imply
 smoother evolution.}
\item[]{{\bf (2)} The best frequency band to observe.
 If we consider all wavelengths, the theory tells us that 
the total intensity is equal to the surface brightness, so the chosen bandwidth
should include most of the SED in the interval $\nu_e$ and $\nu_r$.}
\item[]{{\bf (3)} If the galaxies chosen are located in clusters or are field galaxies.}
\end{enumerate}

\begin{acknowledgements}
I.O.-S. is grateful to CAPES for the financial support.
\end{acknowledgements}

\bibliographystyle{aa}

\end{document}